\documentclass{osa-article}

\journal{oe}

\begin{document}

\title{Stabilization of transmittance fluctuations caused by beam wandering in continuous-variable quantum communication over free-space atmospheric channels}

\author{Vladyslav C. Usenko,\authormark{1,*} Christian Peuntinger,\authormark{2,3} Bettina Heim,\authormark{2,3,4}
Kevin G\"unthner,\authormark{2,3} Ivan Derkach,\authormark{1} Dominique Elser,\authormark{2,3} 
Christoph Marquardt,\authormark{2,3} Radim Filip,\authormark{1} and Gerd Leuchs\authormark{2,3}
}

\address{\authormark{1}Department of Optics, Palack\'y University, 17. listopadu 12, 77146 Olomouc, Czech Republic\\
\authormark{2}Max-Planck-Institut f\"ur die Physik des Lichts, Staudtstr. 2, 91058 Erlangen, Germany\\
\authormark{3}Institut f\"ur Optik, Information und Photonik, Universit\"at Erlangen-N\"urnberg, Staudtstr. 7/B2, 91058 Erlangen, Germany\\
\authormark{4}Currently at OHB System AG, Manfred-Fuchs-Str. 1, 82234 Oberpfaffenhofen, Germany
}

\email{\authormark{*}usenko@optics.upol.cz}


\begin{abstract}
Transmittance fluctuations in turbulent atmospheric channels result in quadrature excess noise which limits applicability of continuous-variable quantum communication. Such fluctuations are commonly caused by beam wandering around the receiving aperture. We study the possibility to stabilize the fluctuations by expanding the beam, and test this channel stabilization in regard of continuous-variable entanglement sharing and quantum key distribution. We perform transmittance measurements of a real free-space atmospheric channel for different beam widths and show that the beam expansion reduces the fluctuations of the channel transmittance by the cost of an increased overall loss. We also theoretically study the possibility to share an entangled state or to establish secure quantum key distribution over the turbulent atmospheric channels with varying beam widths. We show the positive effect of channel stabilization by beam expansion on continuous-variable quantum communication as well as the necessity to optimize the method in order to maximize the secret key rate or the amount of shared entanglement. Being autonomous and not requiring adaptive control of the source and detectors based on characterization of beam wandering, the method of beam expansion can be also combined with other methods aiming at stabilizing the fluctuating free-space atmospheric channels.
\end{abstract}

\section{Introduction}
The development of experimental quantum optics in the past decades led to the emergence and tremendous progress in the field of quantum information, which studies the possibility to store, transmit and process information encoded into quantum states. Quantum communication, a particular application of quantum information processing, is very naturally suggested by the long coherence time and relatively low coupling to the environment which is typical for optical quantum states. This allows one to use quantum states of light for quantum communication, particularly for sharing a quantum resource (such as entanglement) to connect quantum devices, or for quantum key distribution (QKD), aimed at securely distributing random secret keys between two legitimate parties. The methods of QKD are called protocols and were first suggested on the basis of strongly nonclassical systems such as single photons or entangled photon pairs \cite{Gisin2002}. Later the natural use of continuous-variable (CV) \cite{Braunstein2005} quantum states of light was suggested \cite{Ralph1999}. This resulted in the development of CV QKD protocols and methods to produce, characterize and share CV entanglement. 

CV QKD protocols are typically based on the use of Gaussian quadrature-modulated coherent \cite{Grosshans2002,Grosshans2003} or squeezed states \cite{Cerf2001,Madsen2012} of light and homodyne detection at the receiving station. Equivalently, quadrature-entangled states and homodyne detection at both the sending and the receiving stations can be used \cite{Grosshans2003a}. The security of Gaussian CV QKD protocols \cite{Diamanti2015} was shown against general attacks in the asymptotic regime \cite{Leverrier2013} and against collective attacks in the finite-size regime \cite{Leverrier2010,Ruppert2014} based on the optimality of Gaussian attacks \cite{Navascues2006,Garcia2006,Pirandola2008}. This approach allows to broadly study the security of the protocols using covariance matrices, which explicitly characterize Gaussian states of light \cite{Weedbrook2012a}. Gaussian CV QKD protocols were well studied and successfully implemented in long-distance fiber links \cite{Grosshans2003,Jouguet2013,Huang2016}, where the transmittance is typically stable and the added channel excess noise is extremely low. On the other hand, atmospheric quantum channels, which are of utmost importance for long-distance satellite communication \cite{Bedington2017} or free-space terrestrial communication waiving the requirement of necessity of fiber-optical infrastructure, are typically inclined to transmittance fluctuations due to turbulence effects \cite{Berman2006,Semenov2009,Baskov2018}, also affected by weather conditions \cite{Vasylyev2017}. Such transmittance fluctuations (also referred to as channel fading) were analyzed in their impact on applicability of CV quantum communication in the case of atmospheric turbulence \cite{Usenko2012,Heim2014,Hosseinidehaj2015,Bohmann2016} and uniform transmittance fluctuations \cite{Papanastasiou2018}. It was shown that channel fading can be destructive to CV QKD protocols and limit the possibility to share CV entangled states. The main reason for this is that the transmittance fluctuations lead to additional excess noise appearing in the variances of the quadrature measurement results \cite{Usenko2012}. Such fading-related excess noise is proportional to the variance of the transmittance fluctuations and the overall variance of the quadrature distributions in the quantum signal. Therefore in order to allow CV QKD or quantum resource sharing over a fluctuating channel the stabilization of the channel transmittance can be advantageous as a feasible alternative to channel post selection \cite{Usenko2012} or entanglement distillation \cite{Heersink2006,Dong2008}. In the case of mid-range atmospheric optical channels, the transmittance fluctuations are typically caused by beam wandering, when the beam spot is randomly traveling around the receiving aperture \cite{Churnside1990}, in addition to such turbulence effects, as, e.g., scintillation, phase degradation of the wave front, and beam spreading. The transmittance fluctuations caused by beam wandering are then governed, in particular, by the ratio between the beam size and the size of the aperture \cite{Vasylyev2012}. It was suggested that an increase of this ratio would naturally stabilize the channel and make it more suitable for quantum communication tasks \cite{Usenko2012}, similarly to optimization of the beam spot size for given channel parameters in classical free-space optical communication \cite{Guo2010,Ren2010}. 

In the present paper we discuss the method of beam expansion, aimed at compensating the channel fluctuations caused by beam wandering, in detail for CV quantum communication tasks, where the signal intensity is drastically limited compared to the classical free-space optical communication. We report the experimental test of the method based on the spatial expansion of the beam and the subsequent characterization of the channel transmittance. We show that the fading can be indeed stabilized and the variance of transmittance fluctuations (and, subsequently, quadrature excess noise) can be substantially reduced at the cost of increase of the overall loss of the channel. This leads to the trade-off between channel stabilization and its applicability to entanglement sharing or CV QKD. Therefore the suggested method of channel stabilization should be optimized to reach maximum key rate or secure distance for CV QKD or maximum shared entanglement in practical quantum applications.

\section{Fading due to beam wandering in CV quantum communication}
The most feasible CV quantum communication and QKD protocols are based on Gaussian states and operations \cite{Weedbrook2012a}. It is well known that Gaussian states and their properties are explicitly described by the first and the second moments of the field quadrature operators, which can be introduced through the mode's quantum operators as $x=a^\dag+a$ and $p=i(a^\dag-a)$, i.e. by the mean values $\langle x\rangle$, $\langle p \rangle$ and by the covariance matrix $\gamma$ of the elements of the form $\gamma_{i,j}=\langle r_ir_j \rangle-\langle r_i\rangle \langle r_j \rangle$, where $r_i=\{x_i,p_i\}$ is the quadrature vector of the $i$-th mode. It was shown that channel fading leads to excess noise in the quadrature variance, which is proportional to the variance of the channel fluctuations and the variance of the state propagating through the channel \cite{Dong2008} such that the variance of a quadrature on the output of a purely attenuating fading channel becomes $V_{r_i}^{'}=1+\langle \sqrt{\eta}\rangle^2(V_{r_i}-1)+\epsilon_{f_i}$. Here $\langle \sqrt{\eta}\rangle$ is the mean channel transmittance and $\epsilon_{f_i}=Var(\sqrt{\eta})(V_{r_i}-1)$ is the excess noise due to fading, which depends on the variance of the transmittance fluctuations $Var(\sqrt{\eta})=\langle \eta \rangle - \langle \sqrt{\eta} \rangle^2$ and the $r_i$-quadrature variance $V_{r_i}$ of the source. Noise due to fading is therefore generally phase-sensitive, but we further, with no loss of generality, assume phase-space symmetry of the considered states, having variance $V_{r_i}=V : \forall i$ in any quadrature, and subsequent phase independence of the noise $\epsilon_f=Var(\sqrt{\eta})(V-1)$. This noise reduces and possibly destroys the entanglement of a Gaussian state shared over a fading channel and as well decreases the secret key rate of the Gaussian CV QKD. It can lead to loss of security in CV QKD \cite{Usenko2012}, i.e., turning the key rate to zero. The effect is more pronounced for stronger transmittance fluctuations, lower mean channel transmittance and larger initial state variance $V$.

One of the main causes of transmittance fluctuations in an atmospheric channel is beam wandering \cite{Churnside1990},  when the optical beam moves around the aperture of the receiving detector and becomes clipped. It was studied for transmission of quantum states of light for which the transmittance distribution was shown to be governed by the log-negative Weibull distribution, cut at a certain value of transmittance $\eta_0$ \cite{Vasylyev2012,Vasylyev2016}. The distribution is then given by the scale and shape parameters, expressed by the beam-center position variance $\sigma_b^2$ and the ratio $a/W$ of the aperture radius $a$ and the beam-spot radius $W$ so that the maximum transmittance is defined by $\eta_0^2=1-exp[-2(a/W)^2]$. The beam-spot fluctuations variance $\sigma_b^2$ is related to the Rytov parameter \cite{Vasylyev2018}, defining the turbulence strength, which can be obtained from the atmospheric structure constant of refractive index $C_{n}$ \cite{Andrews2001}. The latter, however, was not directly measured in our experiment and further we describe the beam-spot fluctuations by the variance $\sigma_b^2$. It was naturally predicted that the expansion of the beam i.e. the decrease of the ratio $a/W$ would result in a stabilization of the channel transmittance at the cost of a decrease of the mean transmittance \cite{Usenko2012}, a technique also used in classical optical communication \cite{Guo2010,Ren2010}. In our research we verify and confirm this conjecture and study the effect of beam expansion on the channel properties, the efficiency of sharing quantum entanglement and on the security of CV QKD through a fading channel.

\section{Experimental set-up and results}
\begin{figure}[h!]
\centering\includegraphics[width=1.0\textwidth]{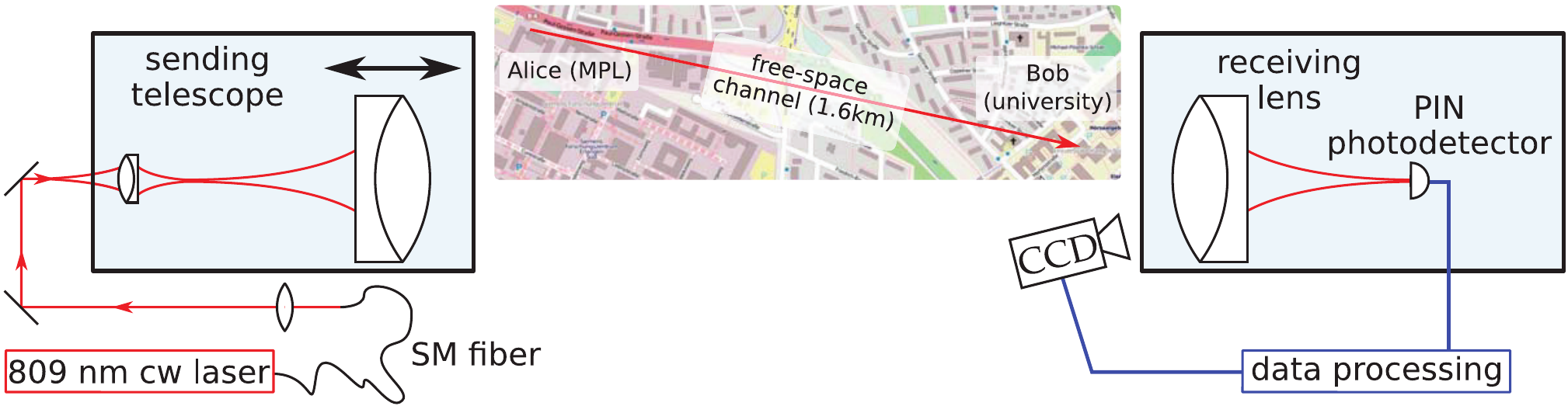}
\caption{A schematic view of the experimental set-up. At the sender Alice we use a telescope to expand the beam and adjust its beam width. Subsequently the beam is sent through our 1.6\,km free-space link to the receiver Bob. There we use an achromatic lens with a diameter of $a=150$\,mm and measure the fluctuating transmission using a PIN photodiode detector and an analogue-to-digital converter. To estimate the aperture-to-beam size ratio we use a CCD camera and a screen.
\label{setup}}
\end{figure}

\begin{figure}[h]
\centering
\includegraphics[width=0.48\textwidth]{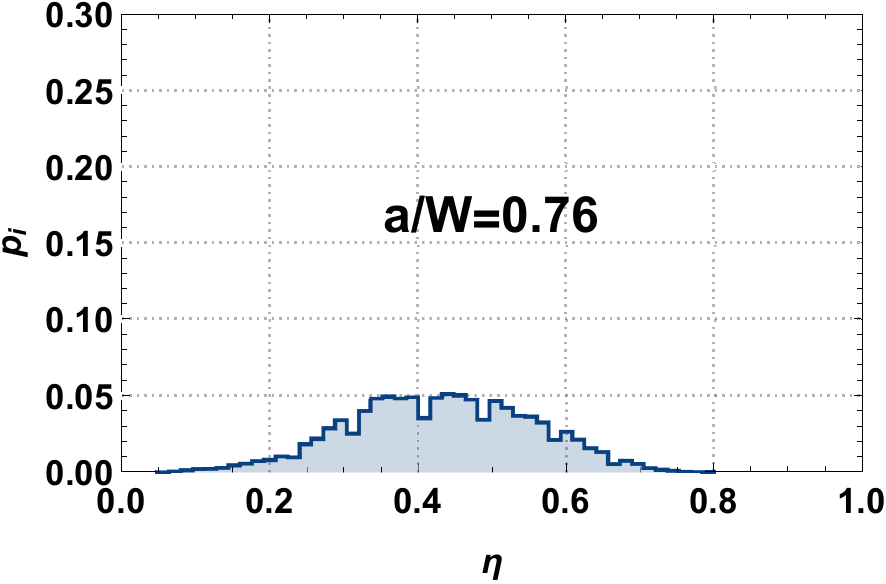}
\quad
\includegraphics[width=0.48\textwidth]{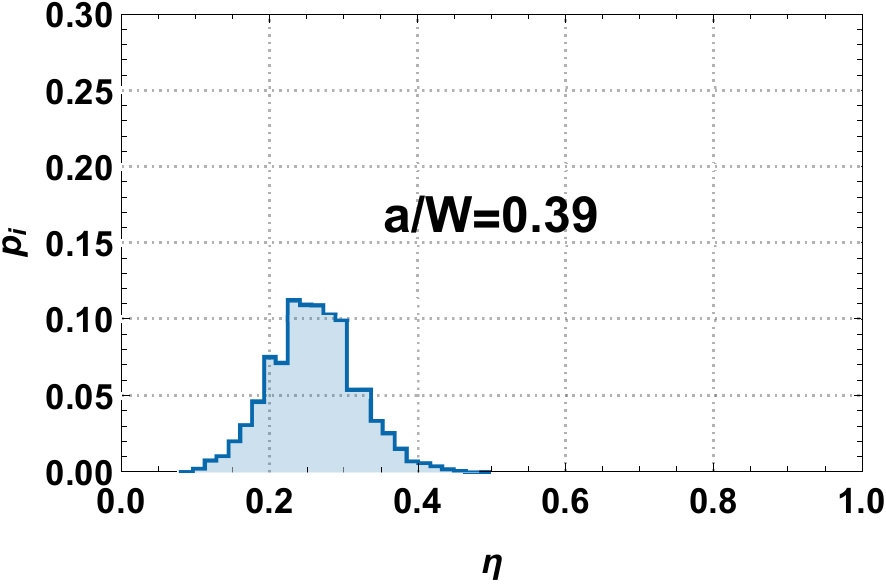}	\\
\includegraphics[width=0.48\textwidth]{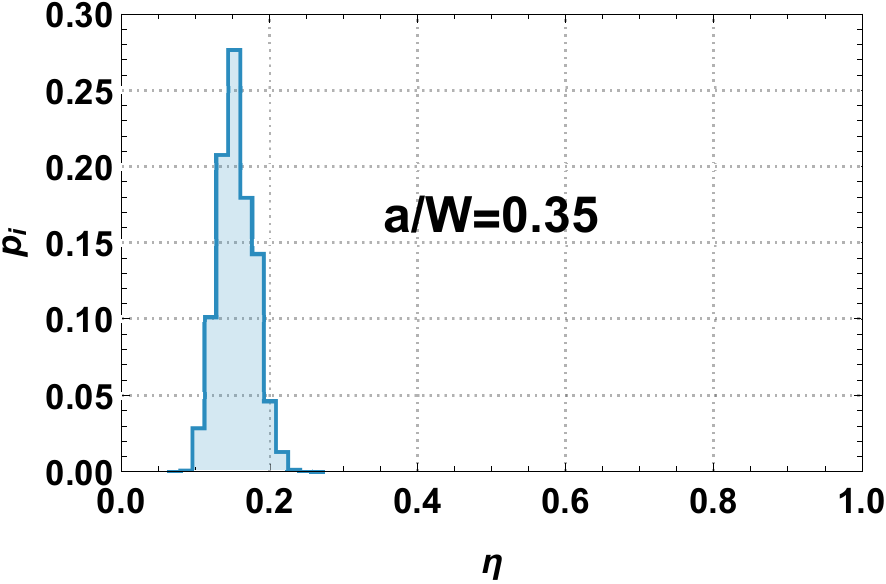}
\quad
\includegraphics[width=0.48\textwidth]{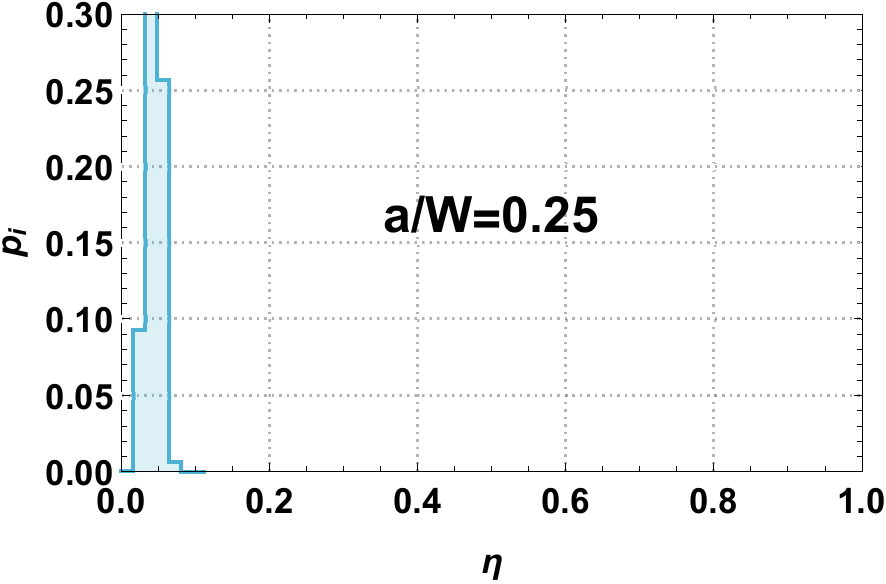}
\caption{Transmittance distribution profiles for different aperture-to-beam size ratios as indicated at the plots.
\label{hists}}
\end{figure}

\begin{figure}[h]
\centering
\begin{tabular}{lll}
\includegraphics[width=0.45\textwidth]{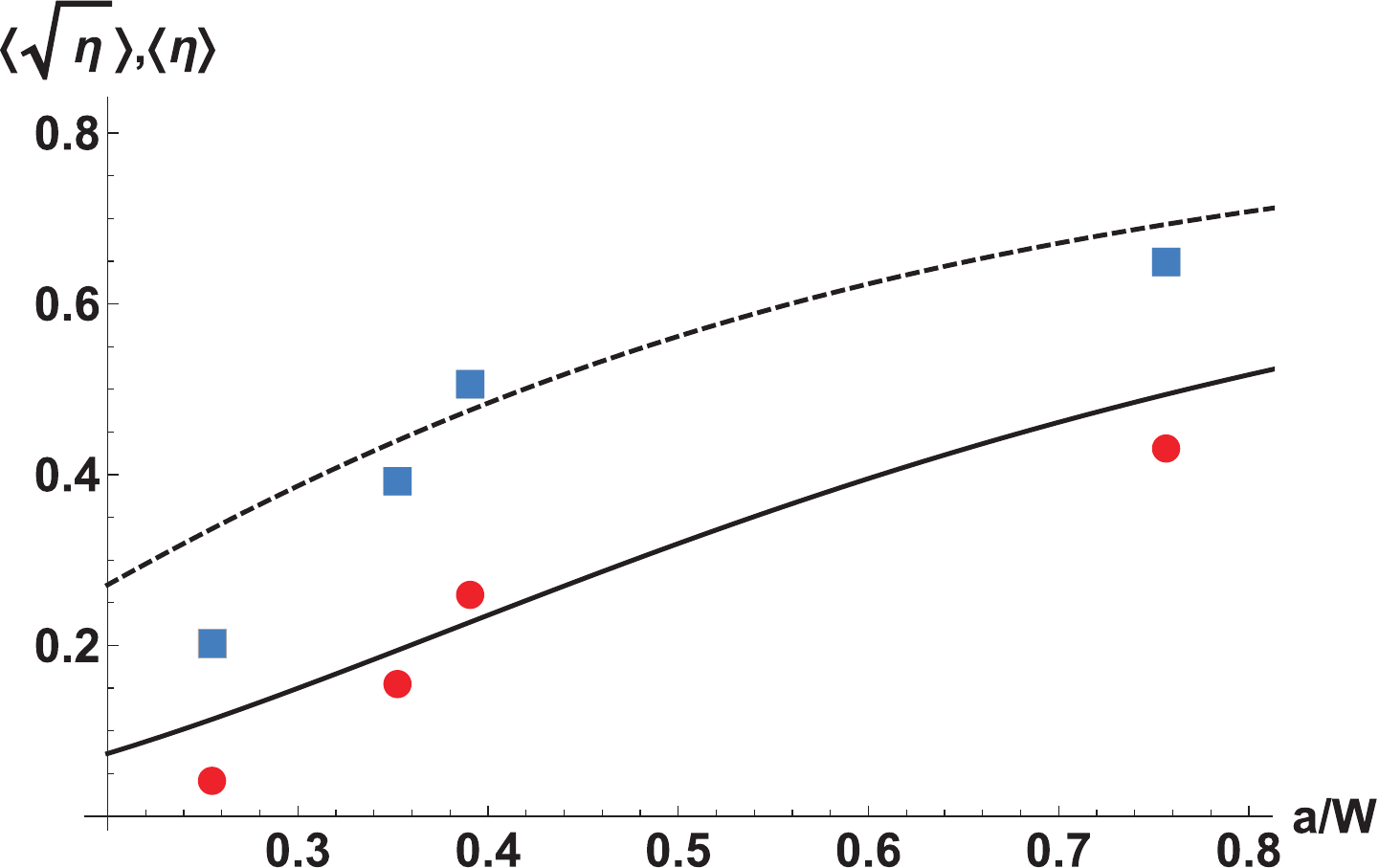}
\quad
\includegraphics[width=0.45\textwidth]{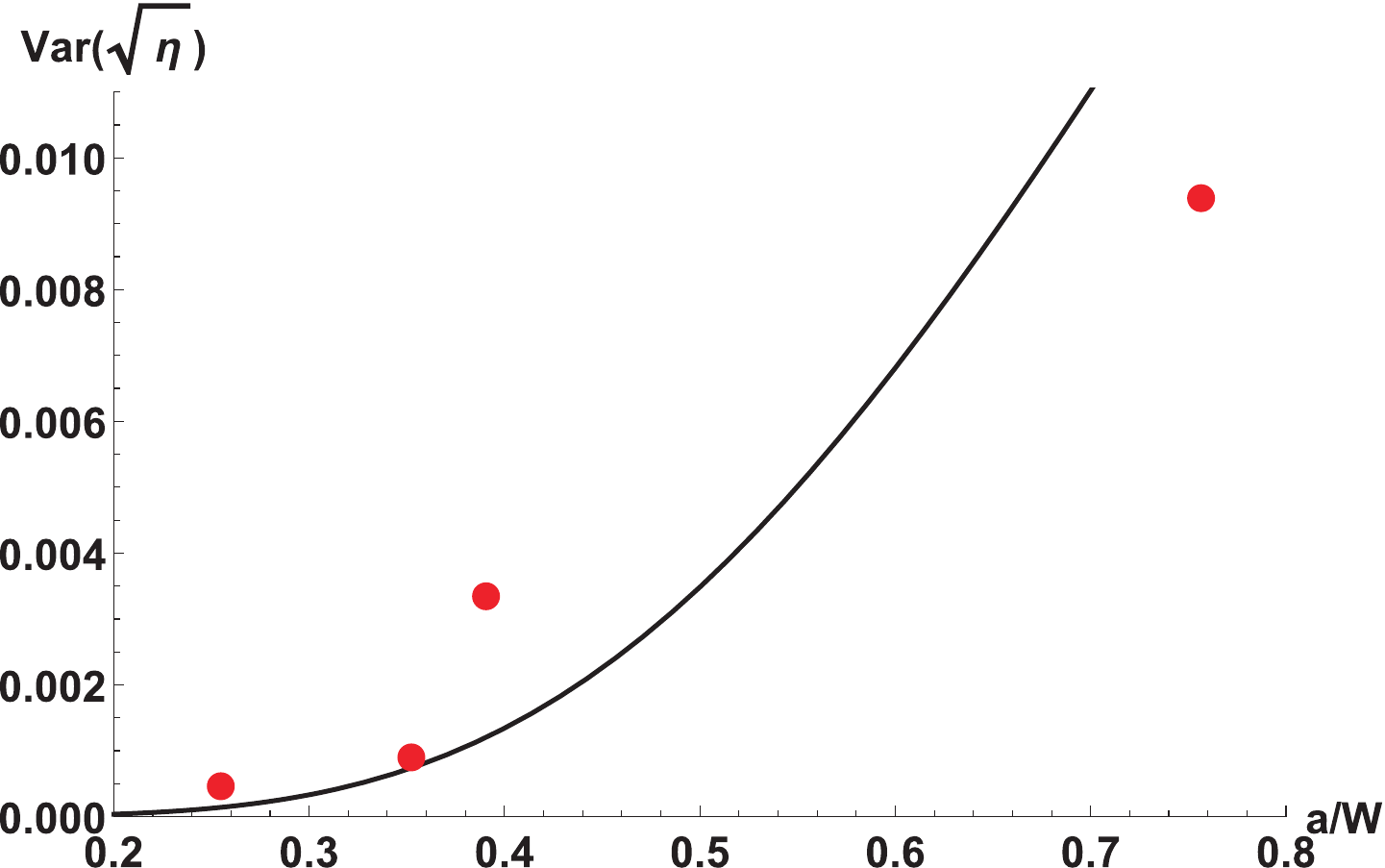}	
\end{tabular}
\caption{Characteristics of atmospheric fading channel for larger beam expansion characterized by decreasing aperture-to-beam size ratio $a/W$. (Left): mean values $\langle \eta\rangle$ (lower solid black line) and $\langle\sqrt{\eta}\rangle$ (upper dashed grey line) estimated from the analytical Weibull distribution along with the experimental results (squares and circles respectively) and (right): variance $Var(\sqrt{\eta})$ of the square root of transmittance from the experimental characterization of the channel (points) and from the analytical estimates (solid line) versus aperture-to-beam size ratio.
\label{means}}
\end{figure}

The possibility to stabilize the fading channel by expanding the beam was studied in a real-world scenario in the city of Erlangen. The used point-to-point free-space channel of 1.6 km length connects the building of the Max Planck Institute for the Science of Light with the building of the computer sciences of the Friedrich-Alexander-University Erlangen-N\"urnberg. We use a grating stabilized continuous wave diode laser with a wavelength of $\lambda=809$\,nm. The mode of this laser is cleaned using a single mode fiber before the beam is expanded using a telescope (see Fig \ref{setup}). The beam is then sent through the fading free-space channel to Bob. At Bob we use an achromatic lens with a diameter of $a=150$\,mm and a focal length of 800\,mm, which defines our aperture. The beam width of the received beam, i.e.\ the aperture-to beam size ratio $a/W$, can be adjusted with the sender telescope. A PIN photodiode detector (bandwidth 150\,kHz) is used to measure the fluctuating transmittance of the channel. To estimate the beam width at Bob we use a CCD camera and a screen.
No adaptive strategy has been used at Bob's station or between Alice and Bob. The profiles of the transmittance distributions for different beam expansion settings that illustrate the change of the statistics of the channel fading are given in Fig. \ref{hists}.
The transmittance data was analyzed to obtain the mean values of transmittance $\langle \eta\rangle$ and $\langle\sqrt{\eta}\rangle$, and the resulting variance $Var(\sqrt{\eta})$, which governs the evolution of a covariance matrix after propagating through the fading channel. The results are given in Fig. \ref{means} along with the values, obtained from the analytical Weibull distribution for the beam-spot fluctuation variance of $\sigma_b^2=0.3$, which is set so in all the subsequent calculations except for these, resulting in the plots in Fig. \ref{KRsigma}. 
%
The results of calculations from the experimentally obtained data demonstrate qualitatively the same tendencies with the decrease of the aperture-to-beam size ratio as the theoretical prediction: it is clearly visible from the plots that the expansion of the beam (i.e., decrease of the aperture-to-beam size ratio) reduces the fluctuations of the transmittance  and at the same time reduces the average transmittance of the channel. In order to clarify the effect of the channel stabilization by the beam expansion on the quantum communication and quantum resource sharing we apply the obtained characteristics of the channel to these applications in the next section.

\section{Effect of beam expansion on entangled resource sharing and CV QKD}
\begin{figure}[h]
\centering
\begin{tabular}{lll}
\includegraphics[width=0.45\textwidth]{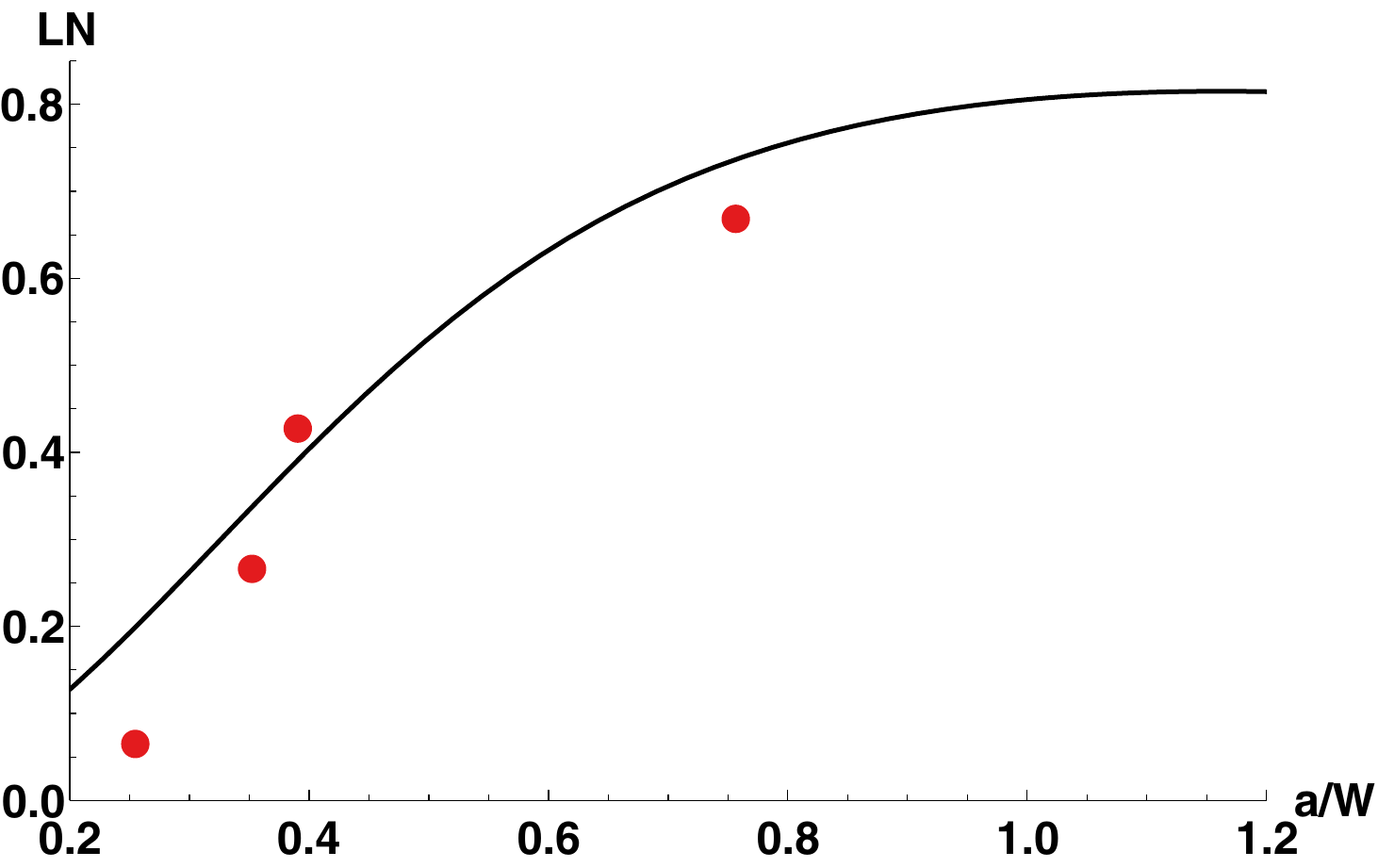}
\quad
\includegraphics[width=0.47\textwidth]{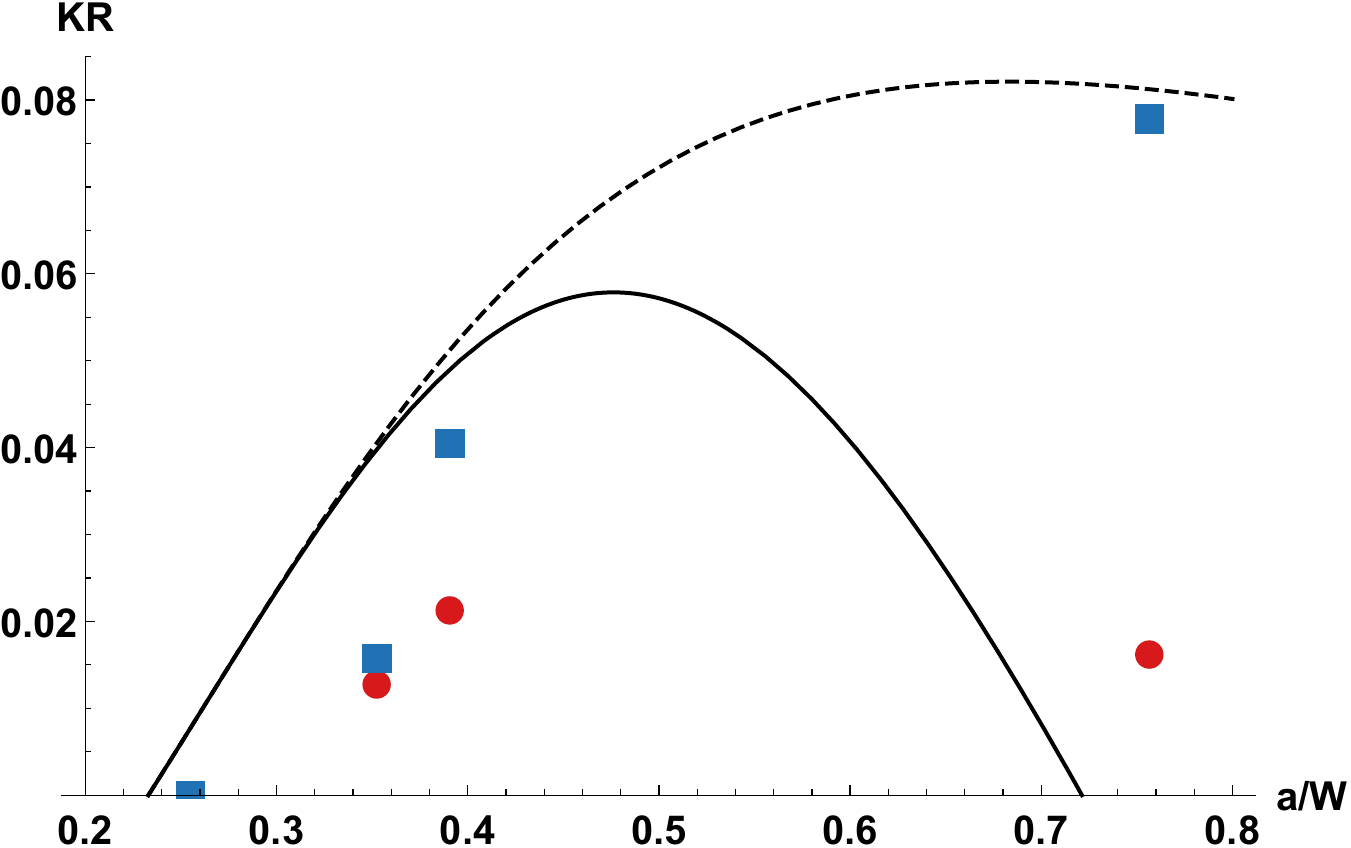}	
\end{tabular}
\caption{Entanglement and secure key rate for larger beam expansion characterized by decreasing aperture-to-beam size ratio a/W. (Left): Logarithmic negativity of an entangled state shared over the fading channel and (Right) Lower bound on the key rate secure against collective attacks  in the fading channel, obtained from the analytical fading distribution (lines) along with the experimental results (points) versus aperture-to-beam size ratio. State variance is 7 SNU (solid black line, red circles) or optimized (dashed black line, blue squares), channel excess noise is 1\% SNU, post-processing efficiency for the Gaussian CV QKD is 97\%.
\label{apps}}
\end{figure}

Before we analyze the applicability of channel stabilization by beam expansion for CV QKD, we first study the impact of the method on the entanglement of a typical two-mode Gaussian entangled state, namely two-mode squeezed vacuum \cite{Weedbrook2012a}, shared over a fading channel. We characterize the entanglement of the state using the logarithmic negativity \cite{Vidal2002}, defined as
\begin{equation}
 LN =\max \lbrace 0, - \log_2 \nu \rbrace ,
\end{equation}
where $\nu$ is the smallest symplectic eigenvalue of a covariance matrix of a partially transposed state for a pair of modes (see \cite{Weedbrook2012a} for review on covariance matrix formalism for Gaussian states). We evaluate the logarithmic negativity for a state quadrature variance of $V=7$ shot-noise units (SNU, being the variance of the vacuum fluctuations), corresponding to approximately -8 dB of conditionally prepared quadrature squeezing after a homodyne detection on one of the beams, which is feasible with current technology \cite{Eberle2013} and is close to optimum for the given protocol parameters, in the presence of 1\% SNU of excess noise (here and further the fixed channel excess noise is related to the channel input). The results of the calculations are given in Fig. \ref{apps} (left) obtained from the experimental data and from the analytical fading distribution. It is clear from the graphs that the channel for the non-expanded beam was more suitable for entanglement distribution and that the beam expansion degraded the entanglement due to increase of the overall loss. The reason for such behavior is that in the considered region of parameters Gaussian entanglement is more sensitive to the channel transmittance than to the small amount of excess noise caused by fading. The transmittance fluctuations in the studied channels were relatively low and did not introduce significant noise, which would reduce the Gaussian entanglement of the states, while decreasing the average transmittance due to beam expansion resulting in entanglement degradation. 

%
%
%

\begin{figure}[h]
\centering
\includegraphics[width=0.6\textwidth]{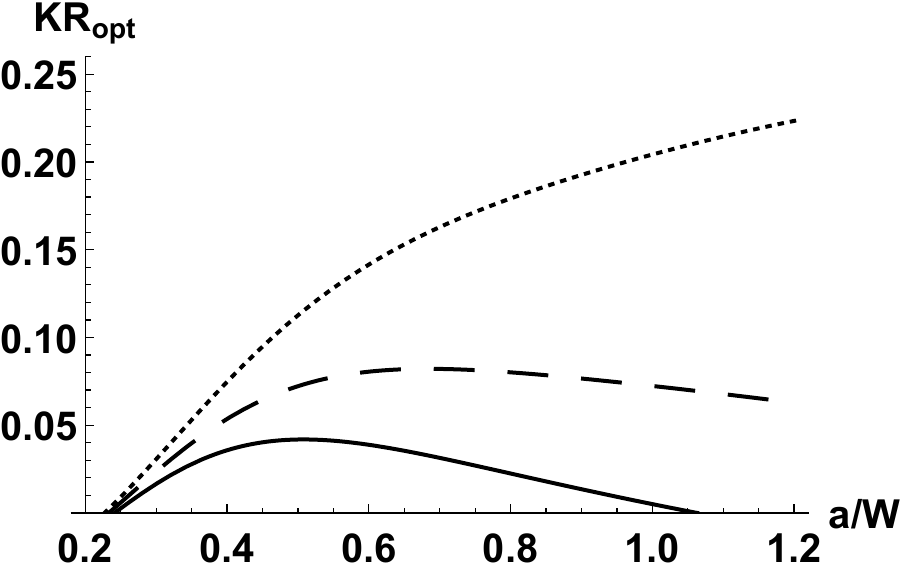}
\caption{Lower bound on the key rate secure against collective attacks  in the fading channel, obtained from the analytical fading distribution, versus aperture-to-beam size ratio at $\sigma_b^2=0.2$ (upper, dotted line),$\sigma_b^2=0.3$ (middle, dashed line), $\sigma_b^2=0.4$ (lower, solid line) upon optimized modulation variance, channel excess noise is 1\% SNU, post-processing efficiency is 97\%.
\label{KRsigma}}
\end{figure}

We also analyze the effect of the beam expansion on the typical CV QKD protocol with coherent states of light and homodyne detection by theoretically estimating the lower bound on the key rate secure against collective attacks \cite{Devetak2005} in a given channel, which, in the reverse reconciliation scenario (being robust against channel attenuation below -3 dB \cite{Grosshans2003}), is given by
\begin{equation}
KR=\beta I_{AB}-\chi_{BE},
\end{equation}
where $I_{AB}$ is the classical (Shannon) mutual information between the trusted parties, $\chi_{BE}$ is the Holevo bound on an information on the shared key received by the remote party, which is available to an eavesdropper, $\beta \in (0,1)$ is the post-processing efficiency, which characterizes how close the trusted parties are able to reach the mutual information $I_{AB}$. In our analysis we follow the purification-based method (see \cite{Usenko2016} for the details of security analysis) to calculate the Holevo bound \cite{Holevo2001} and take into account the realistic post-processing efficiency of 97\% \cite{Jouguet2011}. The results of the calculations are given in Fig. \ref{apps} (right) and clearly show the improvement of the key rate due to the stabilization of the fading channel with a small beam expansion upon fixed modulation, which, however, becomes disadvantageous upon the further increase of the beam spot. We therefore confirm the positive effect of the beam expansion in the fading channel on the CV QKD, which, however, can be optimized in the particular conditions. For Gaussian entanglement distribution or for the coherent-state CV QKD protocol with optimized modulation the method would have been useful for a stronger channel turbulence. The positive role of fading stabilization for the optimized CV QKD upon stronger turbulence is theoretically predicted in Fig. \ref{KRsigma}, where the lower bound on the key rate is plotted versus the beam expansion settings at different values of beam-spot fluctuations. It is evident from the plot, that the experimentally tested beam expansion settings would have been advantageous for the optimized protocol at $\sigma_b^2=0.4$ (note that in our previous study of the same channel upon stronger turbulence the beam-spot fluctuations variance was estimated as $\sigma_b^2=0.36$ \cite{Usenko2012}). 

\begin{figure}[h]
\centering
\begin{tabular}{lll}
\includegraphics[width=0.49\textwidth]{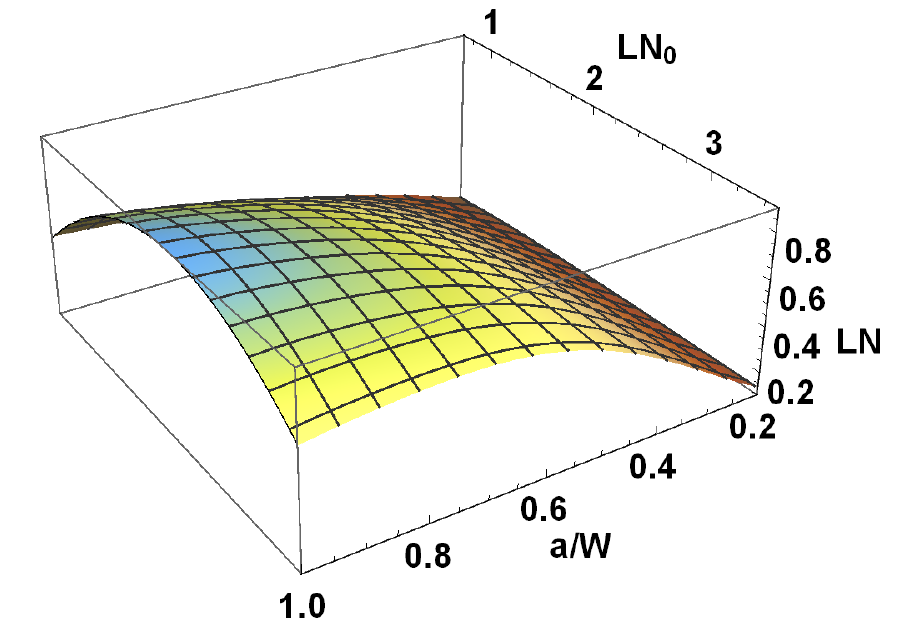}
\quad
\includegraphics[width=0.47\textwidth]{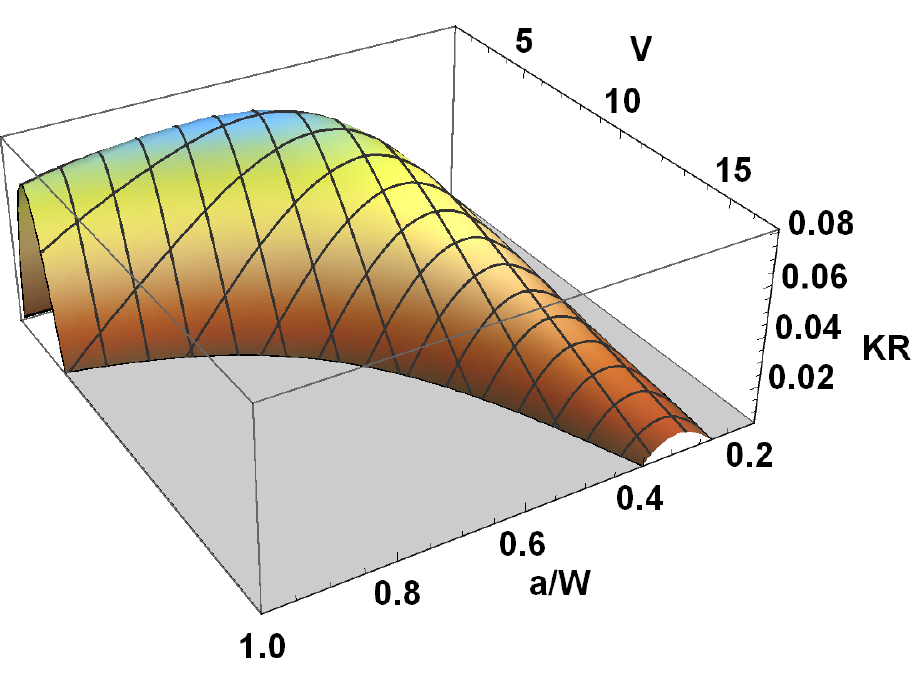}
\end{tabular}
\caption{Entanglement and secure key rate versus aperture-to-beam size ratio a/W at different initial resources. (Left): Logarithmic negativity (LN) of an entangled state shared over a fading channel versus aperture-to-beam size ratio and initial logarithmic negativity $LN_0$ and (Right) Lower bound on the key rate (KR) secure against collective attacks  in the fading channel versus aperture-to-beam size ratio and modulated state variance $V$. Channel excess noise is 1\% SNU, post-processing efficiency for the Gaussian CV QKD is 97\%.
\label{3dapps}}
\end{figure}

Despite evident differences in the effect of beam expansion on the considered quantities (namely logarithmic negativity and key rate) as shown in Fig. \ref{apps}, we theoretically observe a similar behavior of logarithmic negativity at higher values of $a/W$ ratio (out of the experimentally tested and plotted region). Indeed, the logarithmic negativity also has a local maximum at certain ratio $a/W$ (depending on the variance $V$), similarly to the key rate, and would decrease for higher values of the ratio. The difference is however that the key rate is more sensitive to channel fluctuations due to beam wandering and we therefore observed an improvement of the channel parameters by means of beam expansion for the application of CV QKD. Moreover, entanglement is not vanishing completely at high $a/W$ for moderate initial entanglement corresponding to a variance of $V<15$. To complete our study, we numerically illustrate the behavior of the logarithmic negativity and the lower bound on the secure key rate with respect to ratio $a/W$ in the given channel for different initial resources in Fig. \ref{3dapps}. We characterize the initial resource by the state variance $V$ for the key rate plot or, equivalently, by the initial entanglement of the shared state, which reads $LN_0=(-1/2)\ln{(2V^2-1-2V\sqrt{V^2-1})}$ for the logarithmic negativity plot, to verify how much of the initial entanglement survives in a fading channel. It is evident from the plots that beam expansion in the considered channel can have positive effect on the key rate practically for any modulation and on the entanglement once the initial entanglement and beam-to-aperture ratio are large.
In our study we considered the most feasible coherent-state CV QKD protocol. While squeezed-state protocol is known to be typically more robust against channel transmittance fluctuations, its performance is still degraded by fading, related to beam wandering \cite{Derkach2018}, so the beam expansion technique can be useful for the squeezed-state protocols as well and should be optimized in the given conditions.

\section{Conclusions}
We studied the possibility to stabilize a real fading channel by expanding the beam in order to suppress the transmittance fluctuations concerned with the beam wandering in turbulent atmosphere. We experimentally characterized the change of statistics of the channel transmittance fluctuations and showed that they qualitatively correspond to the theoretical predictions given by the Weibull distribution. We proved the positive effect of the channel stabilization by beam expansion on the distribution of a nonclassical resource (entanglement) and on Gaussian continuous-variable quantum key distribution. We have shown that for the channel used for the experimental results presented here beam expansion could become disadvantageous for Gaussian entanglement of the distributed state, described by the logarithmic negativity, due to weak atmospheric turbulence. On the other hand, channel stabilization by beam expansion can improve the secret key rate of the coherent-state protocol. The improvement requires an optimization of the beam width setting under given conditions. Importantly, the method does not require any adaptive control of the source and detector based on monitoring of the beam wandering. It can be combined with other known methods for fading channel stabilization such as fast steering \cite{Suite2004} or concave mirrors \cite{Hulea2014}, channel diversity \cite{Navidpour2007,Lee2004}, multiple wavelengths \cite{Liu2009}, adaptive optics and active tracking systems \cite{Baister1994,Epple2007,ArockiaBazilRaj2016,Kaushal2017,Son2017}. It should be emphasized that our technique requires a link which allows for a certain margin in the loss tolerance. Especially for satellite links the loss is usually already very high as the aperture-to-beam size is very low, such that our stabilization technique will hardly have any benefit. But our proposed technique can be beneficial in mid-range terrestrial free-space links, what would be the field of application for this stabilization technique. Our result therefore demonstrates a promising and feasible method to stabilize free-space atmospheric channels for the tasks of continuous-variable quantum key distribution and quantum communication, which is best applicable in low or medium loss regime. Future steps will include full implementation of continuous-variable quantum key distribution and entanglement sharing over free-space atmospheric channels aided by channel stabilization methods.

\section*{Funding}
Czech Ministry of Education (M\v{S}MT) (LTC17086); EU COST Action (CA15220); Palack\'y University (IGA-PrF-2018-010)

\section*{Acknowledgments}
V.C.U. thanks A.~A.~Semenov for discussions, C.P., B.H. and K.G. thank their colleagues at the FAU computer science building for their kind support and for hosting the receiver.


\begin{thebibliography}{10}
\newcommand{\enquote}[1]{``#1''}

\bibitem{Gisin2002}
N.~Gisin, G.~Ribordy, W.~Tittel, and H.~Zbinden, ``Quantum cryptography,'' Rev. Mod. Phys. \textbf{74}, 145 (2002).

\bibitem{Braunstein2005}
S.~L.~Braunstein and P.~Van Loock, ``Quantum information with continuous variables,'' Rev. Mod. Phys. \textbf{77}, 513
(2005).

\bibitem{Ralph1999}
T.~C.~Ralph, ``Continuous variable quantum cryptography,'' Phys. Rev. A \textbf{61}, 010303 (1999).

\bibitem{Grosshans2002}
F.~Grosshans and P.~Grangier, ``Continuous variable quantum cryptography using coherent states,'' Phys. Rev. Lett.
\textbf{88}, 057902 (2002).

\bibitem{Grosshans2003}
F.~Grosshans, G.~Van Assche, J.~Wenger, R.~Brouri, N.~J.~Cerf, and P.~Grangier, ``Quantum key distribution using
gaussian-modulated coherent states,'' Nature \textbf{421}, 238--241 (2003).

\bibitem{Cerf2001}
N.~J.~Cerf, M.~Levy, and G.~Van Assche, ``Quantum distribution of gaussian keys using squeezed states,'' Phys. Rev. A
\textbf{63}, 052311 (2001).

\bibitem{Madsen2012}
L.~S.~Madsen, V.~C.~Usenko, M.~Lassen, R.~Filip, and U.~L.~Andersen, ``Continuous variable quantum key distribution
with modulated entangled states,'' Nat. Commun. \textbf{3}, 1083 (2012).

\bibitem{Grosshans2003a}
F.~Grosshans, N.~J.~Cerf, J.~Wenger, R.~Tualle-Brouri, and P.~Grangier, ``Virtual entanglement and reconciliation
protocols for quantum cryptography with continuous variables,'' Quantum Inf. Comput. \textbf{3}, 535--552 (2003).

\bibitem{Diamanti2015}
E.~Diamanti and A.~Leverrier, ``Distributing secret keys with quantum continuous variables: Principle, security and
implementations,'' Entropy \textbf{17}, 6072--6092 (2015).

\bibitem{Leverrier2013}
A.~Leverrier, R.~Garc\'{i}a-Patr\'{o}n, R.~Renner, and N.~J.~Cerf, ``Security of continuous-variable quantum key distribution
against general attacks,'' Phys. Rev. Lett. \textbf{110}, 030502 (2013).

\bibitem{Leverrier2010}
A.~Leverrier, F.~Grosshans, and P.~Grangier, ``Finite-size analysis of a continuous-variable quantum key distribution,''
Phys. Rev. A \textbf{81}, 062343 (2010).

\bibitem{Ruppert2014}
L.~Ruppert, V.~C.~Usenko, and R.~Filip, ``Long-distance continuous-variable quantum key distribution with efficient
channel estimation,'' Phys. Rev. A \textbf{90}, 062310 (2014).

\bibitem{Navascues2006}
M.~Navascu\'{e}s, F.~Grosshans, and A.~Ac\'{i}n, ``Optimality of Gaussian attacks in continuous-variable quantum
cryptography,'' Phys. Rev. Lett. \textbf{97}, 190502 (2006).

\bibitem{Garcia2006}
R.~Garc\'{i}a-Patr\'{o}n and N.~J.~Cerf, ``Unconditional optimality of Gaussian attacks against continuous-variable quantum
key distribution,'' Phys. Rev. Lett. \textbf{97}, 190503 (2006).

\bibitem{Pirandola2008}
S.~Pirandola, S.~L.~Braunstein, and S.~Lloyd, ``Characterization of collective Gaussian attacks and security of
coherent-state quantum cryptography,'' Phys. Rev. Lett. \textbf{101} (2008).

\bibitem{Weedbrook2012a}
C.~Weedbrook, S.~Pirandola, R.~Garc\'{i}a-Patr\'{o}n, N.~J.~Cerf, T.~C.~Ralph, J.~H.~Shapiro, and S.~Lloyd, ``Gaussian
quantum information,'' Rev. Mod. Phys. \textbf{84}, 621 (2012).

\bibitem{Jouguet2013}
P.~Jouguet, S.~Kunz-Jacques, A.~Leverrier, P.~Grangier, and E.~Diamanti, ``Experimental demonstration of long-distance
continuous-variable quantum key distribution,'' Nat. Photonics \textbf{7}, 378--381 (2013).

\bibitem{Huang2016}
D.~Huang, P.~Huang, D.~Lin, and G.~Zeng, ``Long-distance continuous-variable quantum key distribution by controlling
excess noise,'' Sci. Rep. \textbf{6}, 19201 (2016).

\bibitem{Bedington2017}
R.~Bedington, J.~M.~Arrazola, and A.~Ling, ``Progress in satellite quantum key distribution,'' npj Quantum Inf. \textbf{3}, 30
(2017).

\bibitem{Berman2006}
G.~Berman and A.~Chumak, ``Photon distribution function for long-distance propagation of partially coherent beams
through the turbulent atmosphere,'' Phys. Rev. A \textbf{74}, 013805 (2006).

\bibitem{Semenov2009}
A.~A.~Semenov and W.~Vogel, ``Quantum light in the turbulent atmosphere,'' Phys. Rev. A \textbf{80}, 021802 (2009).

\bibitem{Baskov2018}
R.~Baskov and O.~Chumak, ``Laser-beam scintillations for weak and moderate turbulence,'' Phys. Rev. A \textbf{97}, 043817
(2018).

\bibitem{Vasylyev2017}
D.~Vasylyev, A.~Semenov, W.~Vogel, K.~G\"{u}nthner, A.~Thurn, \"{O}.~Bayraktar, and C.~Marquardt, ``Free-space quantum
links under diverse weather conditions,'' Phys. Rev. A \textbf{96}, 043856 (2017).

\bibitem{Usenko2012}
V.~C.~Usenko, B.~Heim, C.~Peuntinger, C.~Wittmann, C.~Marquardt, G.~Leuchs, and R.~Filip, ``Entanglement of
gaussian states and the applicability to quantum key distribution over fading channels,'' New J. Phys. \textbf{14}, 093048
(2012).

\bibitem{Heim2014}
B.~Heim, C.~Peuntinger, N.~Killoran, I.~Khan, C.~Wittmann, C.~Marquardt, and G.~Leuchs, ``Atmospheric continuous variable
quantum communication,'' New J. Phys. \textbf{16}, 113018 (2014).

\bibitem{Hosseinidehaj2015}
N.~Hosseinidehaj and R.~Malaney, ``Gaussian entanglement distribution via satellite,'' Phys. Rev. A \textbf{91}, 022304
(2015).

\bibitem{Bohmann2016}
M.~Bohmann, A.~A.~Semenov, J.~Sperling, and W.~Vogel, ``Gaussian entanglement in the turbulent atmosphere,'' Phys.
Rev. A \textbf{94}, 010302 (2016).

\bibitem{Papanastasiou2018}
P.~Papanastasiou, C.~Weedbrook, and S.~Pirandola, ``Continuous-variable quantum key distribution in uniform
fast-fading channels,'' Phys. Rev. A \textbf{97}, 032311 (2018).

\bibitem{Heersink2006}
J.~Heersink, C.~Marquardt, R.~Dong, R.~Filip, S.~Lorenz, G.~Leuchs, and U.~L.~Andersen, ``Distillation of squeezing
from non-gaussian quantum states,'' Phys. Rev. Lett. \textbf{96}, 253601 (2006).

\bibitem{Dong2008}
R.~Dong, M.~Lassen, J.~Heersink, C.~Marquardt, R.~Filip, G.~Leuchs, and U.~L.~Andersen, ``Experimental entanglement
distillation of mesoscopic quantum states,'' Nat. Phys. \textbf{4}, 919 (2008).

\bibitem{Churnside1990}
J.~H.~Churnside and R.~J.~Lataitis, ``Wander of an optical beam in the turbulent atmosphere,'' Appl. Opt. \textbf{29}, 926
(1990).

\bibitem{Vasylyev2012}
D.~Y.~Vasylyev, A.~Semenov, and W.~Vogel, ``Toward global quantum communication: beam wandering preserves
nonclassicality,'' Phys. Rev. Lett. \textbf{108}, 220501 (2012).

\bibitem{Guo2010}
H.~Guo, B.~Luo, Y.~Ren, S.~Zhao, and A.~Dang, ``Influence of beam wander on uplink of ground-to-satellite laser
communication and optimization for transmitter beam radius,'' Opt. Lett. \textbf{35}, 1977--1979 (2010).

\bibitem{Ren2010}
Y.~Ren, A.~Dang, B.~Luo, and H.~Guo, ``Capacities for long-distance free-space optical links under beam wander
effects,'' IEEE Photonics Technol. Lett. \textbf{22}, 1069--1071 (2010).

\bibitem{Vasylyev2016}
D.~Vasylyev, A.~Semenov, and W.~Vogel, ``Atmospheric quantum channels with weak and strong turbulence,'' Phys. Rev. Lett. \textbf{117}, 090501 (2016).

\bibitem{Vasylyev2018}
D.~Vasylyev, W.~Vogel, and A.~Semenov, ``Theory of atmospheric quantum channels based on the law of total
probability,'' Phys. Rev. A \textbf{97}, 063852 (2018).

\bibitem{Andrews2001}
L.~C.~Andrews, R.~L.~Phillips, and C.~Y.~Hopen, Laser beam scintillation with applications, vol. 99 (SPIE press,
2001).

\bibitem{Vidal2002}
G.~Vidal and R.~F.~Werner, ``Computable measure of entanglement,'' Phys. Rev. A \textbf{65}, 032314 (2002).

\bibitem{Eberle2013}
T.~Eberle, V.~H\"{a}ndchen, and R.~Schnabel, ``Stable control of 10 db two-mode squeezed vacuum states of light,'' Opt.
Express \textbf{21}, 11546--11553 (2013).

\bibitem{Devetak2005}
I.~Devetak and A.~Winter, ``Distillation of secret key and entanglement from quantum states,'' Proc. Royal Soc. A:
Math. Phys. Eng. Sci. \textbf{461}, 207--235 (2005).

\bibitem{Usenko2016}
V.~C.~Usenko and R.~Filip, ``Trusted noise in continuous-variable quantum key distribution: A threat and a defense,''
Entropy \textbf{18}, 20 (2016).

\bibitem{Holevo2001}
A.~S.~Holevo and R.~F.~Werner, ``Evaluating capacities of bosonic Gaussian channels,'' Phys. Rev. A \textbf{63}, 032312
(2001).

\bibitem{Jouguet2011}
P.~Jouguet, S.~Kunz-Jacques, and A.~Leverrier, ``Long-distance continuous-variable quantum key distribution with a
Gaussian modulation,'' Phys. Rev. A \textbf{84}, 062317 (2011).

\bibitem{Derkach2018}
I.~Derkach, V.~C.~Usenko, and R.~Filip, ``Squeezing-enhanced quantum key distribution over atmospheric channels,''
arXiv:1809.10167 [quant-ph] (2018).

\bibitem{Suite2004}
M.~R.~Suite, H.~R.~Burris, C.~I.~Moore, M.~J.~Vilcheck, R.~Mahon, C.~Jackson, M.~F.~Stell, M.~A.~Davis, W.~S.
Rabinovich, W.~J.~Scharpf, A.~E.~Reed, and G.~C.~Gilbreath, ``Fast steering mirror implementation for reduction of
focal-spot wander in a long-distance free-space optical communication link,'' in \textit{Free-Space Laser Communication
and Active Laser Illumination III}, D.~G.~Voelz and J.~C.~Ricklin, eds. (SPIE, 2004).

\bibitem{Hulea2014}
M.~Hulea, Z.~Ghassemlooy, S.~Rajbhandari, and X.~Tang, ``Compensating for optical beam scattering and wandering
in fso communications,'' J. Light. Technol. \textbf{32}, 1323--1328 (2014).

\bibitem{Navidpour2007}
S.~M.~Navidpour, M.~Uysal, and M.~Kavehrad, ``Ber performance of free-space optical transmission with spatial
diversity,'' IEEE Trans. Wirel. Commun. \textbf{6}, 2813 (2007).

\bibitem{Lee2004}
E.~J.~Lee and V.~W.~Chan, ``Part 1: Optical communication over the clear turbulent atmospheric channel using
diversity,'' 	IEEE J. Sel. Areas Commun. \textbf{22}, 1896--1906 (2004).

\bibitem{Liu2009}
X.~Liu, ``Free-space optics optimization models for building sway and atmospheric interference using variable
wavelength,'' IEEE Trans. Commun. \textbf{57}, 492--498 (2009).

\bibitem{Baister1994}
G.~Baister and P.~Gatenby, ``Pointing, acquisition and tracking for optical space communications,'' Electron. \&
Commun. Eng. J. \textbf{6}, 271--280 (1994).

\bibitem{Epple2007}
B.~Epple and H.~Henniger, ``Discussion on design aspects for free-space optical communication terminals,'' IEEE
Commun. Mag. \textbf{45}, 62 (2007).

\bibitem{ArockiaBazilRaj2016}
A.~ArockiaBazilRaj and U.~Darusalam, ``Performance improvement of terrestrial free-space optical communications
by mitigating the focal-spot wandering,'' J. Mod. Opt. \textbf{63}, 2339--2347 (2016).

\bibitem{Kaushal2017}
H.~Kaushal and G.~Kaddoum, ``Optical communication in space: Challenges and mitigation techniques,'' IEEE Commun. Surv. Tutor. \textbf{19}, 57--96 (2017).

\bibitem{Son2017}
I.~K.~Son and S.~Mao, ``A survey of free space optical networks,'' Digital Communications and Networks \textbf{3}, 67--77 (2017).

\end{thebibliography}
\end{document}